\documentclass[%
aip,
jap,%
numerical,
amsmath,amssymb,
reprint,%
]{revtex4-1}

\usepackage{graphicx}% Include figure files
\usepackage{dcolumn}% Align table columns on decimal point
\usepackage{bm}% bold math
\usepackage{dcolumn}% Align table columns on decimal point
\usepackage{dsfont}% bold math
\usepackage{bbm}% bold math
\usepackage{hyperref}% add hypertext capabilities
\usepackage[mathlines]{lineno}% Enable numbering of text and display math
\usepackage{natbib}

\begin{document}
\title{Radiative heat transfer between nanoparticles: shape dependence and three-body effect} %Insert here a short version of the title if it exceeds 70 characters

\author{Omid Ramezan Choubdar, Moladad Nikbakht}

\email{mnik@znu.ac.ir}
\affiliation{Department of Physics, University of Zanjan, Zanjan 45371-38791, Iran}

\begin{abstract}
We study the effect of particles shape on the radiative heat transfer in a three-body system. It is found that the radiative heat flux between two nanoparticles in a three body system can be tuned by the shape of the third particle. In particular, we show that the heat flux is very sensitive to the particle shapes and slight mismatches of shapes results in either enhanced or suppressed heat flux.

\end{abstract}

\maketitle

%%%%%%%%%%%%%%%%%%%%%%%%%%%%%%%%%%%%%%%%%%%%%%%%%%%%%%%%%%%%%%%%%%%%%%%%%%%

%%%%%%%%%%%%%%%%%%%%%%%%%%%%%%%%%%%%%%%%%%%%%%%%%%%%%%%%%%%%%%%%%%%%%%%%%%%
\section{Introduction}
The ability to control the radiative heat flux at the nanoscale is the cornerstone for a wide range of applications, including energy conversion, data storage and thermal sensing \cite{swanson,srit,arvind,laro,basut}. It is well-known that the radiative transfer between objects depends drastically on their separation distance. Bringing two objects at small distances changes the radiative heat transfer compared to the classical radiative transfer in the far-field \cite{polder,hu,roussea,shen,ottens,Domingues}. It is shown both theoretically and experimentally that the radiative heat flux spectrum between two object (as emitter and/or absorber), depends on their separation distance and the dielectric constant of the host material \cite{benhost1,Arvind2,Perez,chapuis,ale}. Several techniques are developed to the fabrication of nanostructures and synthesis of nanoparticles with varied size, shape, and composition \cite{fabrication,synthesis1,synthesis2}. It has become increasingly evident that the collective response of nanoparticles, in colloidal suspensions or those deposited on a substrate, exhibits prominent differences in comparison with their isolated response \cite{opticalproperties,fractal,nikbakht3,nikbakht4}. An understanding of the thermal properties of groups of nanoparticles holds both fundamental and practical significance. Fundamentally, it is important to systematically explore nanostructure characteristics that cause radiative property variation. Practically, the tunable radiative properties of nanostructures can be developed into many new application in thermal management \cite{thermalmanagement,solarcell}, thermal optical data storage \cite{thermastorage} and nanofluidics\cite{nanofluid1,nanofluid2,nanofluid3}. Apart from separation distances, the dominant frequencies in the radiative heat transfer depend strongly on the optical response of nanoparticles. It is shown by several authors that the heat flux can drastically be influenced in cases when a mismatch exists between the dielectric function of objects \cite{bor,Manjavacas}. This effect, beside the temperature dependence of objects properties, is widely used in designing thermal memories \cite{kub,dyakov} and thermal diodes \cite{otey, bendiod, yang,liz, basu,Ito,bendi}.

During the past few years significant attention has been paid to manipulate the heat in two-body systems. Increasing the number of objects in a system influences the radiative property due to many-body effects. Each particle in a many-body system, acts as a scatterer of the radiative energy. As a result, the particles’ shapes and dielectric functions are main issues, together with their geometric arrangement in the system \cite{Huth, kruger,ben1,messina,ywang,nikbakht,nikbakht2}.

The majority of recent studies in heat flux between nanoparticles in three-body systems have focused on heat flux modification by relative position, sizes and orientation of particles. In this paper, we focus on a different subset of such systems, in which mismatch exists between shape of particles. We show that a small mismatch of shapes has a giant effect on the radiative heat transfer in a three-body system. A framework we have used in calculating heat transfer, is based on Landauer formalism for N-body system \cite{ben1}, extended to anisotropic nanoparticles \cite{nikbakht}. We have calculated the transmission probability between two particles in the presence of the third particle. Particular concentration is devoted to the expression for energy transmission probability in terms of the third particle polarizability tensor. To this end, we spectrally tune the third particle fundamental electric resonances with respect to the other particles by varying its shape. Numerically calculated heat transmission spectra reveals that the heat flux between objects in a many-body system can be suppressed or enhanced depending on the overlapping of the dipolar resonance of nanoparticles in the system. It is shown that a small mismatch of shapes results in either increase or decrease in the heat exchange between two particles compared to the two-body case.
\begin{figure}
\centering
\includegraphics[]{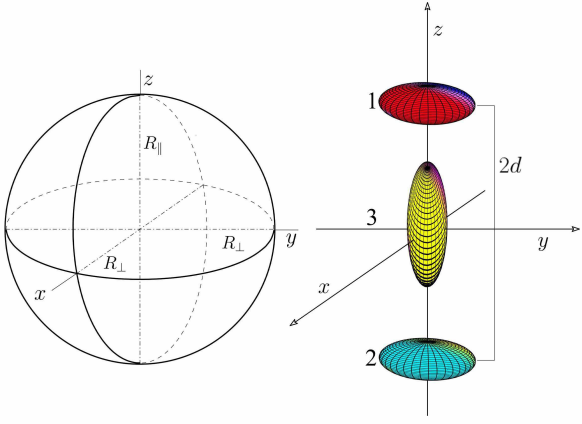}
\caption{ Schematic illustration of the spheroidal nanoparticles in a three-body system. Three spheroids, with identical volumes, are aligned along their axes of symmetry (i.e. $R_\|$) along the z-axis. Particles labeled with 1 and 2 are identical in shape (e.g., prolate spheroid, sphere or oblate spheroid) and located on the z-axis at $\pm d$ at constant temperatures $T_1\neq T_2$. The third particle with varying shape (which may vary from prolate to an oblate spheroid) is located at $z=0$ and used to tune the heat exchange between the two other particles. }\label{fig1}
\end{figure}

\section{Basic Equations}
Let us start by describing the geometrical setup of our physical system. The system considered in our study is an array of three nanoparticles along the z axis of the cartesian coordinate as shown in Fig. \ref{fig1}. Particles which are labeled with 1 and 2 are separated by $2d$ apart and the third nanoparticle (labeled with 3) is placed between them to form a three-body system. Particles 1 and 2 are supposed to be held at different temperatures, $T_1$ and $T_2$, and exchange energy through radiation in the presence of the third particle. The net heat flux between particles 1 and 2 is calculated and it is shown that the heat flux can be tunes with the shape of the third nanoparticle. For the sake of simplicity, we focus our study on homogeneous spheroidal nanoparticles. The electric dipole resonances of spheroidal particles are very sensitive to their shape and can be tuned by the aspect ratio of their main axes. The principal axes of particles are along the main axes of the cartesian coordinate as shown in Fig. \ref{fig1}. The ratio of the semiaxis $R_\|$ (parallel to the z direction) to the semiaxis $R_\bot$ (perpendicular to z direction) characterizes the shape of each particle, which may vary from a nearly spherical ($R_\|/R_\bot \simeq 1$) to a nanodisk ($R_\|/ R_\bot\ll1$) or a nanorod one ($R_\|/ R_\bot\gg1$). Here, we assume that nanoparticles are small enough compared to the smallest separation distance and thermal wavelength $\lambda_{th}=c\hbar/(k_BT)$ in the system (c is the vacuum light velocity, $2\pi\hbar$ is Planck's constant, and $k_B$ is Boltzmann's constant). This assumption insures that the heat transfer between particles can be calculated through simple fluctuating dipoles. The net radiative heat exchanged between two particles ({\it i}th and {\it j}th) in a N-body system can be written in the general form \cite{nikbakht}

\begin{equation}\label{eq.2}
{\mathcal H}_{ij}=\int_0^\infty\frac{d\omega}{2\pi}{\mathcal T}_{ij}(\omega)\Delta\Theta (\omega)
\end{equation}
where $\Delta\Theta(\omega)=\Theta(\omega,T_i)-\Theta(\omega,T_j)$ is the difference of mean energies of Planck oscillators at frequency $\omega$ and at the temperatures of two interacting nanoparticles. Here, ${\mathcal T}_{ij}(\omega)$ represents the energy transmission probability between {\it i}th and {\it j}th nanoparticles in the presence of all scatterers in a system and is defined as \cite{nikbakht}
\begin{equation}\label{eq.3}
{\mathcal T}_{ij}=2{\tt Im}{\mathrm {Tr}}[{\hat{\mathbb A}}_{ij}{\tt Im}({\hat {\bm \chi}}_{j}){\hat{\mathbb C}}_{ij}^\dag]
\end{equation}
where $\hat{\bm \chi_j}={\hat{\bm \alpha_j}}+k^2{\hat{\bm \alpha_j}}{\hat{\bm G}}_0^\dag{\hat{\bm \alpha_j}}^\dag$ is the susceptibility tensor of the {\it j}th particle with polarizability ${\hat{\bm \alpha}}_j$, $k=\omega/c$ and ${\hat{\bf G}}_0=i\frac{k}{6\pi}\mathbbm{1}$. The non-negatively defined ${\tt Im}{\hat {\bm \chi}}$ guarantees a correct direction for heat flux between particles, i.e. from hotter to colder particle. Moreover, the influence of the geometrical arrangement of nanoparticles as well as the many-body effects on the heat flux are taken into account by $3\times 3$ matrices $\hat{\mathbb A}_{ij}$ and ${\hat {\mathbb C}}_{ij}$ \cite{nikbakht}: 
\begin{subequations}
\begin{eqnarray}
&&\hat{\mathbb A}_{ij}=[\mathbbm{1}-k^2\widetilde{\alpha}\hat{\mathbb W}]^{-1}_{ij}\\
&&\hat{\mathbb C}_{ij}=[k^2\mathbbm{G}\mathbb A]_{ij}
\end{eqnarray}
\end{subequations}
where $\mathbbm{1}$ stands for the unit dyadic tensor, $\hat{\mathbb G}_{ij}=\hat{\bf G}_{ij}$, $\hat{\mathbb W}_{ij}=(1-\delta_{ij})\hat{\bf G}_{ij}$, $\widetilde{\alpha}=diag\{\hat{\bm \alpha}_1,\hat{\bm\alpha}_2,\cdots,\hat{\bm\alpha}_N\}$, and the free space dyadic Green's tensor is:

\begin{eqnarray}
&&\hat{\bf G}_{ij}=\frac{k}{4\pi}\Big[f(kr_{ij})\mathbbm{1}+g(kr_{ij})\frac{{\bf r}_{ij}\otimes{\bf r}_{ij}}{r_{ij}^2}\Big]\\
&&\nonumber f(x)=[x^{-1}+ix^{-2}-x^{-3}]\exp(ix)\\
&&\nonumber g(x)=[-x^{-1}-3ix^{-2}+3x^{-3}]\exp(ix)
\end{eqnarray}
which has the contribution of near-, intermediate- and far-zone terms, $\propto r^{-3}$, $r^{-2}$ and $r^{-1}$ respectively.  Here, ${\bf r}_{ij}$ is the vector linking particles located at points ${\bf r}_{i}$ and ${\bf r}_{j}$ with $r_{ij}=|{\bf r}_{i}-{\bf r}_{j}|$.

For the special case of heat exchange between two objects (labeled with 1 and 2) in the presence of the third object (labeled with 3), Eq. (\ref{eq.3}) reduces to

\begin{widetext}
\begin{equation}\label{eq.4}
\begin{split}
\centering
\displaystyle
{\mathcal T}_{12}^{3-body}(\omega)=2k^4{\mathrm {Tr}}\Biggl\{ \frac{\Big({\hat{\bf G}}_{12}{\hat{\bf G}}_{21}^\dag+2k^2{\tt Re}({\hat{\bf G}}_{12}^\dag{\hat{\bf G}}_{13} {\hat{\bf G}}_{23}{\hat {\bm \alpha}}_{3})
+k^4{\hat{\bf G}}_{13}{\hat{\bf G}}_{31}^\dag{\hat{\bf G}}_{23}^\dag{\hat{\bf G}}_{23}{\hat{\bm \alpha}}_{3}{\hat{\bm \alpha}}_{3}^\dag\Big) {\tt Im}{\hat {\bm \chi}}_1{\tt Im}{\hat {\bm \chi}}_2}{{\hat{\mathbb M}}{\hat{\mathbb M}}^\dag}\Biggl\} 
\end{split}
\end{equation}
with
\begin{equation}\label{eq.5}
{\mathbb M}=\Big(\mathbbm{1}-k^4{\hat{\bm \alpha}}_1{\hat{\bf G}}_{12}{\hat{\bm \alpha}}_2{\hat{\bf G}}_{21}-k^4{\hat{\bm \alpha}}_1{\hat{\bf G}}_{13}{\hat{\bm \alpha}}_3{\hat{\bf G}}_{31}-k^4{\hat{\bm \alpha}}_2{\hat{\bf G}}_{23}{\hat{\bm \alpha}}_3{\hat{\bf G}}_{32}-2k^6{\hat{\bm \alpha}}_1{\hat{\bf G}}_{12}{\hat{\bm \alpha}}_2{\hat{\bf G}}_{23}{\hat{\bm \alpha}}_3{\hat{\bf G}}_{31}\Big)
\end{equation}

\end{widetext}

In the absense of the third particle (i.e., ${\hat{\bm \alpha}}_3\rightarrow 0$ and/or ${\hat{\bf G}}_{13},{\hat{\bf G}}_{23}\rightarrow 0$) Eq. (\ref{eq.4}) reduces to

\begin{equation}\label{eq.6}
{\mathcal T}_{12}^{2-body}(\omega)=2k^4{\mathrm {Tr}}\Biggl\{\frac{\Big( {\hat{\bf G}}_{12} {\hat{\bf G}}_{21}^\dag\Big){\tt Im}{\hat {\bm \chi}}_1{\tt Im}{\hat {\bm \chi}}_2}{{\hat{\mathbb M}}{\hat{\mathbb M}}^\dag}\Biggl\}
\end{equation}
with
\begin{equation}\label{eq.7}
{\mathbb M}=\Big(\mathbbm {1}-k^4{\hat{\bm \alpha}}_1{\hat{\bf G}}_{12}{\hat{\bm \alpha}}_2{\hat{\bf G}}_{21}\Big)
\end{equation}

Equations (\ref{eq.4}-\ref{eq.7}) contain complete information on the transmission probability dependence on nanoparticle characteristics and geometric configuration. It can be seen that the imaginary part of the susceptibility tensor of particles 1 and 2 (corresponding to their absorption) are presented in the transmission probability of both systems. We factorized this term to highlight the influence of (particles 1 and 2) shapes on the transmission probabilities. The first term in the parenthesis of the right hand in Eq.(\ref{eq.4}) corresponds to the direct heat flux between particles 1 and 2 in a three body system. The two other terms represent the indirect (but forward) heat flux between these particles tunneled (scattered/absorbed) through the third particle. To a first approximation, the sum of these two terms can be considered as {\it three-body} effect, specially in case of large separation distances. On the other side, only a term related to the direct heat transfer between particles 1 and 2 appeared in the heat flux of a two-body system (see the numerator of Eq. (\ref{eq.6})). Furthermore, The multiple scattering (i.e. forward and backward) of radiating waves between particles are accounted for by the Fabry-P\`erot like denominator (${\hat{\mathbb M}}$) in the heat transmission probabilities of both two- and three-body systems. While this quantity depends only on the arrangement and characteristics of particles 1 and 2 in a two-body system, it also depends on position and characteristics of the third particles in three-body system.

The transmission probability can show resonance due to the Fabry-P\`erot like denominator appeared in Eqs. (\ref{eq.4}) and (\ref{eq.6}) for three- and two-body systems, respectively. These modes will came into resonance whenever $\mathrm {det}({\hat{\mathcal D}})=0$ inwhich ${\hat{\mathcal D}}={\hat{\mathbb M}}{\hat{\mathbb M}}^\dag$. The resonant eigenfrequencies (eigen modes) satisfy ${\mathcal D}_n=0$ where ${\mathcal D}_n$'s are eigenfunctions of matrix ${\hat{\mathcal D}}$. For the special case of identical spherical nanoparticles, one can show that this condition can be written as ${\tt Re}(1/\alpha)+w_n=0$, where $w_n$'s are the eigenfunctions of the interaction matrix $\hat{\mathbb W}$. For an arbitrary collection of $N$ interacting nanoparticles with volume $v=(4\pi/3)R^3$, there exist at most $3N$ number of such modes (in dipolar limit $kR\ll kd$, where $d$ is the smallest center-center separation distance in the system) that contribute to the resultant heat exchange between particles in the system. The resonance frequencies of these modes can be tuned by varying shape, arrangement, size, and composition of nanoparticles (for more details on resonance in large sphere dimers see \cite{khandekar}). 

While any symmetry in the system (including geometric arrangement and/or shapes) my accompanied by increasing degeneracy of the resonance frequencies, the number of {\it distinct} modes that can be excited are depending sensitively on $d/R$. For periodic arrangement of nanoparticles (or even in random gas of particles), the interaction matrix is highly degenerate. At intermediate ($2R\ll d\sim \lambda_{th}$) to large ($2R\ll \lambda_{th}\ll d$) separation distances, the resonance frequencies in this system congregate to the resonance of susceptibility of an individual particles (surface modes), occurring for a spherical particle at $\epsilon=-2\epsilon_h$ (where, $\epsilon_h$ and $\epsilon$ are the complex dielectric function of the background medium and particle respectively). This property is shown by Ben-Abdallah and his co-workers in sphere-dimer and sphere-trimer systems\cite{ben1}.

It should be emphasized that the contribution of the third particle to the heat flux in a three-body system appeared in both the numerator and denominator of the transmission probability in Eq. (\ref{eq.4}). As maintained before, while the former represents the contribution of the third particle in forward heat flux, the latter affects the resonance frequencies of transmission probability through multiple scattering of radiation. The crucial point to be noted is that the weight with which a mode contributes to the resultant transmission probability depends on the numerator of the transmission probability. 

The weight of the direct heat flux has contribution of far-field regime, $d\gg \lambda_{th}$, where $(R/d)^2$-term dominates, as well as the near field regime, where $(R/d)^6$ is largest. Moreover, the contribution of three-body part leads to $(R/d)^{12}$ to $(R/d)^{3}$ terms due to forward heat flux through the third particle, where in combination with multiple scattering, it decays even faster at small separation distances. In the case of $d\sim 2R$, the dipole approximation is not valid and multipolar modes contribution to the heat transfer should be taken into account. At sufficiently large separation distances (in practice, a distance on the order of few radii in case of spherical particles), the heat transmission probability is mainly decided by the susceptibility of particles 1 and 2 as emitter and/or absorber. This dependence appeared as ${\tt Im}{\hat {\bm \chi}}_1{\tt Im}{\hat {\bm \chi}}_2$ in both two- and three-body systems, which can be optimized by tuning the overlap of polarizability tensor components even by shape or material. As shown by Incardone {\it et.al,} the heat transfer between two spheroids can be switched on/of by changing their relative orientation \cite{incardone}. 
The presence of additional particle leads to significant modification in the transmission probability which depends in a complicated manner on shapes in the limit of small separations. In the case of spherical nanoparticles, the heat flux between two particles can be enhanced by inserting a third nanoparticle in the middle \cite{ben1} and can be tuned by it's size \cite{ywang}.

The comparison between the heat transmission probabilities in three- and two-body systems shows that the heat flux between particles 1 and 2 depends on the absorption and scattering properties of the third particle as well as its position. The {\it many-body} part of the transmission probability, which is related to the presence of the third particle, can be written as
\begin{equation}\label{eq.8}
{\mathcal T}_{12}^{(m)}(\omega)={\mathcal T}_{12}^{3-body}(\omega)-{\mathcal T}_{12}^{2-body}(\omega).
\end{equation}
It is evident from Eqs. (\ref{eq.4})-(\ref{eq.7}) that the major contribution to ${\mathcal T}_{12}^{(m)}$ comes from the indirect heat transfer through the third particle. While the last term in Eq. (\ref{eq.4}), has a positive contribution to the heat flux, the second term has a crucial rule in reduction/enhancement of heat flux in a three-body system. However, the contribution of the last term is small compared to other terms for distances of few radii $d\gtrsim 6R$, where the near-field interaction is still valid. It is therefore legitimate to drop last term,  expand ${\mathbb M}^{-1}$ and keep only the lowest-order terms in Eqs. (\ref{eq.4}) and (\ref{eq.6}). By collecting terms first order in ${\hat{\bm \alpha}}_3$ we obtain
\begin{eqnarray}\label{eq.9}
\nonumber{\mathcal T}_{12}^{(m)}&\propto&{\mathrm {Tr}}\Bigg \{k^6{\tt Re}({\hat{\bm \alpha}}_3){\tt Im}{\hat {\bm \chi}}_1{\tt Im}{\hat {\bm \chi}}_2\Big( 4{\tt Re}[{\hat{\bf G}}_{12}^\dag{\hat{\bf G}}_{13} {\hat{\bf G}}_{23}]\\
&&~~~~~~~+2k^2{\tt Re}[{\hat{\bm \alpha}}_1{\hat{\bf G}}_{13} {\hat{\bf G}}_{13}{\hat{\bf G}}_{12}{\hat{\bf G}}_{21}^\dag]\\
\nonumber&&~~~~~~~+2k^2{\tt Re}[{\hat{\bm \alpha}}_2{\hat{\bf G}}_{23} {\hat{\bf G}}_{23}{\hat{\bf G}}_{12}{\hat{\bf G}}_{21}^\dag]\\
\nonumber&&~~~~~~~+8k^4 {\tt Re}[{\hat{\bf G}}_{12}^\dag{\hat{\bf G}}_{13} {\hat{\bf G}}_{23}] {\tt Re}[{\hat{\bm \alpha}}_{1}{\hat{\bf G}}_{12} {\hat{\bm \alpha}}_{2}{\hat{\bf G}}_{12}]\Big) \\
\nonumber&&~~-k^6{\tt Im}({\hat{\bm \alpha}}_3){\tt Im}{\hat {\bm \chi}}_1{\tt Im}{\hat {\bm \chi}}_2\Big( 4{\tt Im}[{\hat{\bf G}}_{12}^\dag{\hat{\bf G}}_{13} {\hat{\bf G}}_{23}]\\
\nonumber&&~~~~~~~+2k^2{\tt Im}[{\hat{\bm \alpha}}_1{\hat{\bf G}}_{13} {\hat{\bf G}}_{13}{\hat{\bf G}}_{12}{\hat{\bf G}}_{21}^\dag]\\
\nonumber&&~~~~~~~+2k^2{\tt Im}[{\hat{\bm \alpha}}_2{\hat{\bf G}}_{23} {\hat{\bf G}}_{23}{\hat{\bf G}}_{12}{\hat{\bf G}}_{21}^\dag]\\
\nonumber&&~~~~~~~+8k^4 {\tt Im}[{\hat{\bf G}}_{12}^\dag{\hat{\bf G}}_{13} {\hat{\bf G}}_{23}] {\tt Re}[{\hat{\bm \alpha}}_{1}{\hat{\bf G}}_{12} {\hat{\bm \alpha}}_{2}{\hat{\bf G}}_{12}]\Big)\Bigg\}
\end{eqnarray}

Note that Eq. (\ref{eq.9}) reflects a simplified approach, since contributions from the higher order scattering at small separation distances are not shown. It can be seen that, although the heat flux between particles 1 and 2 explicitly depends on the imaginary part of ${\hat{\bm \alpha}}_3$, it depends on the real part of the polarizability tensor as well. The real part of the polarizability, denoted by ${\tt Re}{\hat{\bm \alpha}}_3$ is the dispersive part of the polarizability. Likewise, the imaginary part of the polarizability, denoted by ${\tt Im}{\hat{\bm \alpha}}_3$, is the absorptive part \cite{bonin}. The first term in Eq. (\ref{eq.9}) can be attributed to the third particle contribution to the heat transfer by scattering, and the second term corresponds to the absorption by the third particle. We should notice that in the region near SFM resonance, the real part of polarizability components is anti-symmetric about the resonances while the imaginary part is symmetric. The particular form of ${\mathcal T}_{12}^{(m)}$ depends on the particular arrangement in which particles disposed in the system. Thus we can conclude that, in addition to the geometrical arrangement, the albelo of the third particle (i.e., the competition between its scattering and absorption) has a major rule in the heat transfer in three-body system. 

Since the formalism of the model based on the dipole approximation, the results are valid for arbitrary shape (polarizability tensor), size and distances as long as we stay in this regime. If particle sizes are too large compared to separation distances, the contribution of multipolar interaction is needed for accurate results. However, the dipole limit would be enough to reveals cryptic aspects of heat transfer dependency on shape in these systems. In our analysis, we denote the complex polarizability tensor by $\hat{\bm\alpha}=({\alpha}_{\bot} , {\alpha}_{\|})$, where for convenience we labeled the polarizability parallel to the z-axis as $\alpha_{\|}(=\alpha_z)$, and the polarizability perpendicular to the z-axis as $\alpha_\bot (=\alpha_x=\alpha_y)$. These components are related to the dielectric function of the object by the relation \cite{nov}
\begin{subequations}
\begin{eqnarray}
&&\alpha_\bot=v\frac{\epsilon-\epsilon_h}{\epsilon_h+L_\bot(\epsilon-\epsilon_h)}\label{eq.10a}\\
&&\alpha_\|=v\frac{\epsilon-\epsilon_h}{\epsilon_h+L_\|(\epsilon-\epsilon_h)}\label{eq.10b}
\end{eqnarray}
\label{}
\end{subequations}
where $v=\frac{4\pi}{3}R_\bot^2R_\|$ is the volume of particle and $L_{\bot}$ and $L_\|$ ($2L_\bot+L_\|=1$) are depolarization factors. For an arbitrary ratio of the semiaxes $R_\bot$ and $R_\|$, depolarization factors can be computed from integrals:
\begin{subequations}
\begin{eqnarray}
&&L_\bot=\frac{R_\bot^2R_\|}{2}\int_0^\infty \frac{ds}{({R_\bot^2+s})^2\sqrt{(R_\|^2+s)}}\label{eq.11a}\\
&&L_\|=\frac{R_\bot^2R_\|}{2}\int_0^\infty \frac{ds}{({R_\|^2+s})^2\sqrt{(R_\bot^2+s)}}\label{eq.11b}
\end{eqnarray}
\end{subequations}

By considering a constant volume for particles, the ratio of the paralell depolarization $L_\|$ to the perpendicular depolarization $L_\bot$ (i.e., $\gamma=L_\|/L_\bot$) characterizes the nanoparticle shape, which may vary from a nearly
nano-rod ($L_\bot\rightarrow \frac{1}{2}, L_\|\rightarrow0: \gamma\ll 1$) to a spherical ($L_\bot=L_\|=\frac{1}{3}: \gamma=1$) or to a nanodisk ($L_\bot\rightarrow 0, L_\|\rightarrow 1: \gamma\gg 1$). The components of polarizability tensor in terms of $\gamma$ would be
\begin{subequations}
\begin{eqnarray}
&&\alpha_\bot=V\frac{(\epsilon-1)(\gamma+2)}{(1+\gamma+\epsilon)}\label{eq.12a}\\
&&\alpha_\|=V\frac{(\epsilon-1)(\gamma+2)}{(2+\gamma\epsilon)}\label{eq.12b}
\end{eqnarray}
\end{subequations}
where we set $\epsilon_h=1$.

It is clear from symmetry reason that the product ${\tt Im}{\hat {\bm \chi}}_1{\tt Im}{\hat {\bm \chi}}_2$ in Eq. (\ref{eq.6}) gives a maximum transmission probability in the frequency of the surface phonon modes (SFMs) for a two-body system, in which $\gamma_1=\gamma_2$ and $\epsilon_1=\epsilon_2$. The resonance condition is fulfilled when $\epsilon(\omega_{\bot}^{sr})\simeq -(1+\gamma)$ for $\alpha_\bot$ and $\epsilon(\omega_{\|}^{sr}) \simeq -2/\gamma$ for $\alpha_\|$, implying intense scattering/absorption.  A similar argument holds for the case of three-body system, however, the presence of the third particle allows for a large freedom of tunability. Depending on the shape of the third particle, the heat exchange between particles 1 and 2 may increase or decrease with respect to the two-body case. Besides the electric dipole radiation, the magnetic dipole radiation would contribute to the heat exchange, which is not considered here and may come into resonance at well-defined values of ratio of size that alter the heat transfer.

\begin{figure}
\centering
\includegraphics[]{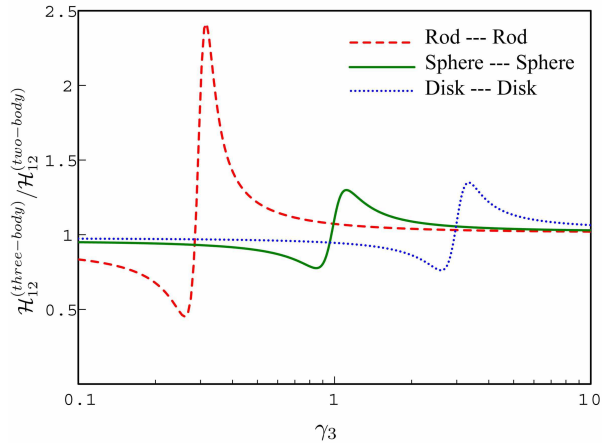}
\caption{Normalized heat exchange between two identical particles with respect to the shape of a third one which is equidistant to both particles. The Rod-Rod (red dashed lines) configuration corresponds to heat exchange between two nanorods with $\gamma_1=\gamma_2=0.3$, the Sphere-Sphere (grin solid line) for two spheres with $\gamma_1=\gamma_2=1$, and Disk-Disk (blue doted line) for two nanodisks with $\gamma_1=\gamma_2=3$. The shape of the third nanoparticle varies from a nearly nano-rod ($\gamma=0.1$) to a nanodiskc($\gamma=10$).}
\label{fig2}.
\end{figure}
\section{Numerical results and discussion}
Let us now, apply these theoretical frameworks to study the shape-dependent heat transfer in a three-body system. In order to reduce the number of parameters, we will focus on the symmetric case in which particles 1 and 2 are identical (i.e., $\gamma_1=\gamma_2$) with separation distance $2d=450$~nm. The shape of the third elliptic particle (i.e., $\gamma_3$) can be used for tuning its scattering properties and thus for tuning the heat exchange between 1 and 2. The volume of nanoparticles are kept constant at that of sphere with $R_\bot=R_\|=25$ nm and we use SiC as a typical material for nano-particles with frequency-dependent relative dielectric permittivity\cite{karl}:

\begin{equation}
\varepsilon(\omega)=\varepsilon_\infty\bigg(1+\frac{\omega_L^2-\omega_T^2}{\omega_T^2-\omega^2-i\Gamma\omega}\bigg)
\label{eq.13}
\end{equation}
with $\omega_L=969$ cm$^{-1}$, $\omega_T=793$ cm$^{-1}$, $\Gamma=4.76$ cm$^{-1}$ and $\varepsilon_\infty=6.7$. Furthermore, we assume that $T_1=300$~K and $T_2=T_3=0$~K, so the first nanoparticle acts as emitter and the heat exchange between particles 1 and 2 reduces to the heat flux from 1 to 2.
The calculation performed for three special cases in which particles 1 and 2 are nanorod, sphere and nanodisk. Figure (\ref{fig2}), represents the normalized heat exchange between particles 1 and 2 with respect to the shape of the third one. 
We have used $\gamma_1=\gamma_2=0.3$ for {\it Rod-Rod}, $\gamma_1=\gamma_2=3$ for {\it Disk-Disk}, and $\gamma_1=\gamma_2=1$ for {\it Sphere-Sphere} configurations. Since $\gamma_1=\gamma_2$, and consequently $\alpha_{\bot,\|}^{(1)}=\alpha_{\bot,\|2}$, the optical properties of the third nanoparticle is very important and plays a vital rule on the heat transfer.

We observe that the heat exchange between particles 1 and 2 depends on the third particle depolarization aspect ratio $\gamma_3$ and may increase or decrease in comparison with the two-body case. When the mismatch of shapes (i.e., $\delta\gamma=|\gamma_3-\gamma_{1,2}|$) is large, the many-body effect is weak, whereas for small mismatches the many-body effect becomes pronounced. For $\gamma_3\gtrsim\gamma_{1,2}$, many-body effect dominated by scattering which gives rise to the heat transfer. In contrast, for $\gamma_3\lesssim\gamma_{1,2}$ absorption is stronger than scattering and as a result the heat exchange decrease in comparison with the two-body system. The maximum (minimum) of the heat transfer profile corresponds to the maximum (minimum) albedo of the third particle around the SFM resonance frequencies of the other particles. On the other hand, the behaviors of the real and imaginary parts of the third particles' polarizability have a crucial rule in many-body effect. This effect is especially strong when the third particle polarizability resonance frequencies are close to the SFM resonance of emitter/absorber particle.

\begin{figure}
\centering
\includegraphics[]{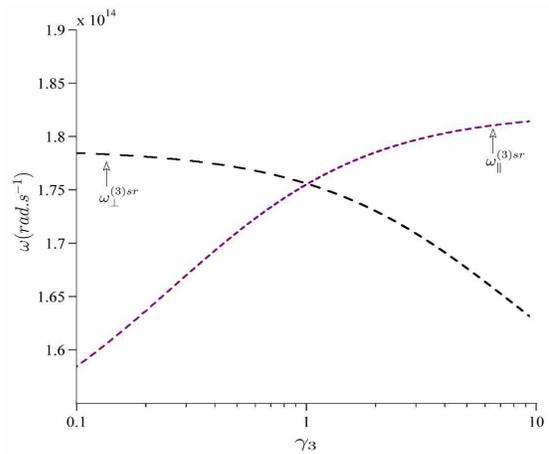}
\caption{Surface resonance frequency components (longitudinal $||$ and transverse $\bot$ to the axis of symmetry) of a SiC spheroid nanoparticle as a function of shape parameter $\gamma$. The particle volume is kept constant at that of sphere with radius $R=25$~nm. The shape factor for a spherical nanoparticle is $\gamma=1$, while $\gamma<1$ and $\gamma>1$ correspond to prolate and oblate spheroids, respectively. The condition for frequency resonances are ${\tt Re}[\epsilon(\omega_{||})]=-2/\gamma$ and ${\tt Re}[\epsilon(\omega_{\bot})]=-(1+\gamma)$ for longitudinal and transverse modes, respectively. }
\label{fig3}.
\end{figure}

\begin{figure*}[t]
%\hrule
%This figure uses both columns, using \texttt{figure*}
\begin{center}
\IfFileExists{graphicx.sty}{
\includegraphics[]{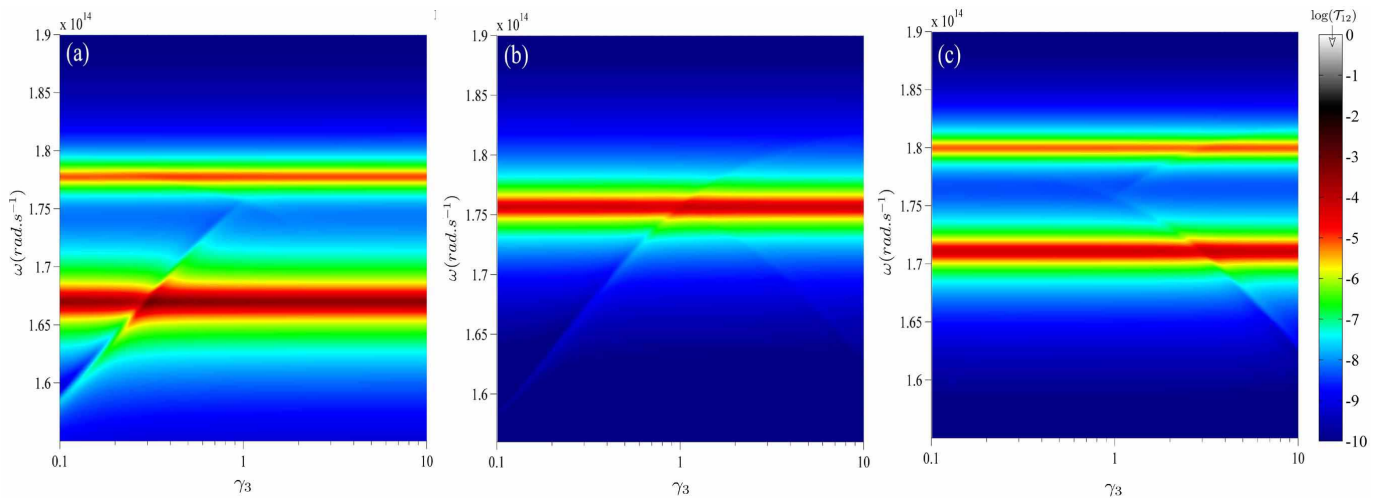}
}{
Sorry, package \texttt{graphicx} not present.
}
\end{center}
\caption{Heat flux spectrum between two identical Spheroid particles with respect to the shape ($\gamma$) of a third one located between them. (a) Particles 1 and 2 in Rod-Rod configuration with $\gamma_1=\gamma_2=0.3$, (b) Particles 1 and 2 in Sphere-Sphere configuration with $\gamma_1=\gamma_2=1.0$, and (c) Particles 1 and 2 in Disk-Disk configuration with $\gamma_1=\gamma_2=3.0$. The horizonal patterns in the spectrums correspond to the resonance surface frequencies of particles 1 and 2, while the cross-like pattern correspond to the resonance surface frequencies of the third particle.
}
\label{fig4}.
\end{figure*}
The frequencies of SFMs are specified by the depolarization factors and subsequently by the shape of nanoparticles. The shape-dependent prolate-to-oblate resonance of the third particles' polarizability components are illustrated in Fig. (\ref{fig3}). Since prolate and oblate nanoparticles have a symmetry axes, they have three proper modes, which two of them are degenerate. As the aspect ratio of the third nanoparticle deviate from one, the energy separation between SFMs increases. In the case of nanorod (i.e., $\gamma_3<1$), the high-energy absorbing band corresponds to the oscillation of the electrons along the $R_\bot$ of the rod. This absorption band is relatively insensitive to the nanorod aspect ratio and spectrally is near to the resonance of the spherical nanoparticle. The other absorption band at lower energies is caused by the oscillation of charges along the $R_\|$ of the rod. As $\gamma_3$ approaches 1 from below, these bands move toward each other and coalesce for spherical nanoparticle. Finally, when $\gamma_3>1$, the switch of energy takes place between the two energy bands. On the other hand, for $\gamma_3>1$, the high-energy absorbing band corresponds to the oscillation of the electrons along $R_\|$, while the {\it two fold} degenerate resonance of the electron oscillation along $R_\bot$ occurs at lower energies for nanodisk.

Depending on the shapes of particles 1 (and 2), the albedo of the third particle affects the energy transmission probability defined in Eq. (\ref{eq.4}). In order to see the physics behind this dependence, the transmission probability spectra ${\mathcal T}_{12}(\omega)$ in a three-body system for a frequency range around the SFMs of the third particle are shown in Figs.~(\ref{fig4}a)-(\ref{fig4}c). Figure (\ref{fig4}a) shows the results for {\it Rod-Rod} configuration, while the transmission spectra for {\it Sphere-Sphere} and {\it Disk-Disk} configurations are presented in Figs. (\ref{fig4}b) and (\ref{fig4}c), respectively. It can be observed that the heat transmission probability shows two main distinct resonance for {\it Rod-Rod} and {\it Disk-Disk} configuration, which reduces to {\it three fold} degenerate SFM in the case of {\it Sphere-Sphere} configuration. Since particles 1 and 2 are identical, the overlap of ${\tt Im}\chi_1$ and ${\tt Im}\chi_2$ is large in all cases. Consequently, the many-body part of the transmission probability mainly depends on the overlap of the third nanoparticle polarizability with the two others. In the case of {\it Rod-Rod} configuration, this overlap occurs for $\gamma_3\approx 0.3$, where both the real and imaginary parts of $\alpha_3$ are in resonance with SFMs of the emitter.
It has to be noted that $\alpha_{3\bot}$ and $\alpha_{3\|}$ show resonance at different frequencies. While the contribution of the longitudinal mode (i. e., $\omega_{\|}^{sr}$) to the three-body effect is pronounced in {\it Rod-Rod} configuration, the contribution of the transverse mode (i. e., $\omega_{\bot}^{sr}$) is dominant in {\it Disk-Disk} configuration. In the case of the {\it Sphere-Sphere} configuration, it can be observed that the heat transmission probability is mediated by the shape of the third particle for polarizability ratio $\gamma_3\approx 1$. By slightly deviation of the third particles' shape from sphere, the heat transmission would increase or decrease in comparison with the two-body case. In the case of {\it Disk-Disk} configuration, the scattering/absorption by the third particle modifies the transmission intensity for $\gamma_3\approx 3$. The resonance of the scattering and absorption by the third particle is located at $\gamma_3\gtrsim 3$ and $\gamma_3\lesssim 3$, respectively. Once again, the change in the transmission probability occurs mainly in low frequencies due to resonance of $\alpha_{3\bot}$. Finally, it should be noticed that in all cases, the third particle's scattering/absorption property is off-resonance for large $\delta\gamma$, and transmission spectra is not affected by the shape of the third particle.
\section{Conclusion} 
In conclusion, we have investigated the effect of the particles' shape in radiative heat transfer. It is shown that the energy transmission probability between two nanoparticles in a three-body system may be increased or decreased, depending on dipolar resonance frequency modes of the third nanoparticle. Also, we have shown the possibility of sensitively tuning the heat flux by particles' shape in such a thermal heat transistor.

\section{References} 

%\bibliography{manuscript1}


\end{document}